\documentclass[a4paper,11pt]{article}
\pdfoutput=1
\usepackage{jheppub}
\usepackage{amsmath,amssymb}
\usepackage{bbold}
\usepackage{graphicx}
\usepackage{float}
\usepackage{color}
\usepackage [utf8]{inputenc}
\usepackage{subfigure}
\usepackage{euscript} 
\usepackage{placeins}
\usepackage{diagbox}
\usepackage[thinlines]{easytable}
\usepackage{epstopdf}

\allowdisplaybreaks[2] 

\def\OO{\mathcal{O}}
\def\bb{{\boldsymbol b}}
\def\bop{{\boldsymbol p}}

\def\alphas{\alpha_{\mathrm{s}}}
\def\gs{g_{\mathrm{s}}}

\def\gsqb{g_{3\mathrm{d}}^2 b_\perp}

\def\gsq{g_{3\mathrm{d}}^2}
\def\gfour{g_{3\mathrm{d}}^4}
\def\gsix{g_{3\mathrm{d}}^6}

\def\mD{m_{\mathrm{D}}}
\def\mDsq{m_{\mathrm{D}}^2}

\def\bp{\bb_\perp}
\def\bbp{b_\perp}
\def\bps{b_\perp^2}

\def\qp{q_\perp}
\def\bqp{\boldsymbol q_\perp}

\def\Cbp{C(\bbp)}
\def\Cqp{C(\qp)}
\def\CQCDbp{C_\mathrm{QCD}(\bbp)}
\def\CEQCDbp{C_\mathrm{EQCD}(\bbp)}

\def\CR{C_\mathrm{R}}
\def\CA{C_\mathrm{A}}
\def\CF{C_\mathrm{F}}

\def\Nc{N_\mathrm{c}}
\def\d{\mathrm{d}}

\def\bmu{\bar{\mu}}
\def\nf{N_{\mathrm{f}}}
\def\Tf{T_{\mathrm{f}}}
\def\qhn{\hat{q}_0}

\newcommand{\qt}{q_\perp}
\newcommand{\qtt}{\mathbf{q}_\perp}
\newcommand{\ptt}{\mathbf{p}_{\perp}}

\newcommand{\btt}{\mathbf{b}}

\newcommand{\zb}{\bar{z}}

\newcommand{\mut}{\tilde{\mu}}

\title{Non-perturbative determination of collisional broadening and medium induced radiation in QCD plasmas}
\author[a]{Guy D.\ Moore} 
\author[b]{S\"oren Schlichting}
\author[a,c]{Niels Schlusser}
\author[b]{Ismail Soudi}
\affiliation[a]{Institut f\"ur Kernphysik, Technische Universit\"at Darmstadt\\
Schlossgartenstra{\ss}e 2, D-64289 Darmstadt, Germany}
\affiliation[b]{Fakultät für Physik, Universität Bielefeld\\ 
D-33615 Bielefeld, Germany}
\affiliation[c]{Department of Physics \& Helsinki Institute of Physics\\
P.O. Box 64, FI-00014 University of Helsinki}
\emailAdd{guy.moore@physik.tu-darmstadt.de, \\
sschlichting@physik.uni-bielefeld.de, \\ 
nschlusser@theorie.ikp.physik.tu-darmstadt.de, \\
isma@physik.uni-bielefeld.de}

\abstract{
We supply recently obtained results from lattice EQCD with the correct UV limit
to construct the collisional broadening kernel $C(\bbp)$ in a QCD plasma. We discuss the limiting behavior of $C(\bbp)$ at small and large impact parameters $\bbp$, and illustrate how the results can be used to compute medium-induced radiation rates.
}


\begin{document}

\maketitle

\section{Introduction}
\label{sec:intro}


An important signal of quark gluon plasma formation during heavy-ion collisions is the suppression of highly energetic particles while they traverse the medium. These highly energetic partons then generate jets,
and it is the suppression of jets, and modifications in jet properties,
which we hope to use as a tool for understanding the medium produced in heavy ion collisions.

The theory community is in some agreement about the process for jet energy loss.
A high-energy particle traversing the medium undergoes a series of soft
scatterings which exchange transverse momentum with the medium.
(Here and throughout, ``transverse" means transverse with respect to
the high energy particle's propagation direction.)
The particle is constantly emitting virtual collinear radiation;
scattering of the particle or the emitted radiation with the medium
can force the radiation to become real.  There are important coherence
effects in this process, originally recognized by Landau and Pomeranchuk
\cite{LP1,LP2} and by Migdal \cite{Migdal1}.
These were explicated within QCD by Zakharov
\cite{Zakharov:1996fv,Zakharov:1997uu,Zakharov:1998sv}
and by Baier et al \cite{Baier:1996kr,Baier:1996sk}.

The physical picture which emerges from this analysis does not appear
to be in doubt.  As a high-energy particle propagates through a medium
with a density matrix $|p\rangle \langle p|$,
a hard vertex generates an amplitude for the mixed state
$|p\rangle \langle p-k,k|$.  This state forms at transverse separation
$\bbp \hspace{-3pt} = 0$ but as time evolves,
it undergoes eikonalized propagation in the
transverse plane, diffusing in $\bp$ but also receiving damping
due to medium interactions; the strength of this damping at transverse
separation $\bbp$, $\Cbp$, is the Fourier transform of
the rate of transverse momentum broadening:
\begin{align}
    \Cqp & \equiv \frac{(2\pi)^2 \d^3 \Gamma}{
    \d^2 \qp \, \d L}
    \,,
    \\
    \Cbp & \equiv \int \! \frac{\d^2 \qp}{(2\pi)^2}
    \left( 1 - e^{i\bqp \cdot \bp} \right) \Cqp \label{subtraction_FT} \, .
\end{align}
So $\Cqp$ is the rate per unit path length and $\bqp$ range to exchange
transverse momentum $\bqp$ with the medium, and $\Cbp$ is its
zero-subtracted Fourier transform.
A subsequent hard vertex can convert this mixed state to a real emission.
Almost all literature treatments are based on this framework, see
Ref.~\cite{Baier:2000mf}.
However, different treatments 
\cite{Gyulassy:1999zd,Gyulassy:2000er,Gyulassy:2003mc}
\cite{Wiedemann:2000za,Salgado:2003gb}
\cite{Arnold:2002ja}
\cite{Djordjevic:2008iz}
\cite{CaronHuot:2010bp}
differ dramatically in their simplifying assumptions and their
treatment of the medium interactions, see
\cite{Armesto:2011ht}.

Treatments also differ in their description of the medium.
One common approximation is to treat the medium as
many random, static, screened color centers:
$\Cqp \propto \frac{1}{(\qp^2 + \mD^2)^2}$.
Dynamical moving charges, treated to lowest order in perturbation
theory, are not much more complicated \cite{Aurenche:2002pd}:
$\Cqp \propto \frac{1}{\qp^2(\qp^2+\mD^2)}$.
It is also common to make the approximation of many individually
small scatterings, leading to transverse momentum diffusion:
$\Cbp = \hat{q} \, \bps/4$.
But these treatments are either assumptions, models,
or low-order perturbative calculations for a medium
which is strongly coupled and where the behavior could have large
nonperturbative contributions even at quite large temperatures.
It would be better to have a treatment of the jet-medium interaction
based more firmly in QCD and with the possibility of including some
genuinely nonperturbative physics.

Recently, this possibility came much closer to reality.
Already more than 10 years ago, Casalderry-Solana and Teaney
showed how the collision kernel $\Cbp$ can be rigorously
defined in terms of the behavior of certain null Wilson loops
\cite{CasalderreySolana:2007qw}, and Caron-Huot showed
how such null Wilson loops could be recast, for temperatures
well above $T_c$, in terms of modified Wilson loops in the
dimensionally reduced long-distance effective theory for QCD,
3D EQCD (3D QCD with adjoint scalars) \cite{CaronHuot:2008ni}.
This theory can be solved nonperturbatively on the lattice,
allowing for the first time for genuine nonperturbative input
into the form of the jet-medium interaction.
In addition, after some important preliminary work
\cite{Panero:2013pla,DOnofrio:2014mld,Moore:2019lua},
we recently presented detailed and continuum-extrapolated results
for $C_{\rm EQCD}(\bbp)$ the impact-parameter space jet-medium interaction
rate within the theory of EQCD \cite{Moore:2020wvy}.

The purpose of this paper is to take this result and to show
how it can be applied in a calculation of jet radiation.
There are two steps which are needed.
First, the work of Ref.~\cite{Moore:2020wvy} is within EQCD,
not within 3+1 dimensional real-time QCD.  We need to complete
the matching between the theories, with the help of some work
by Arnold and Xiao \cite{Arnold:2008vd},
to extract $\Cbp$ for QCD.
We do so in Section \ref{sec:nonpert_Cbp}.
Second, the lattice data for $\Cbp$ covers a finite range
of separations, exists only at discrete $\bbp$ values, and has
errors.  We need to connect it together with the asymptotic small
and large separation limits into a function which can really be
applied in a calculation.  
Then, in Section \ref{sec:inelastic_rates},
we illustrate how to use the results in a real calculation of
jet modification through medium interaction.  For convenience we
carry this out within the AMY formalism
\cite{Arnold:2002ja,Jeon:2003gi,Schenke:2009gb} in a large medium; we will return to its application
in a finite-length medium in a future publication.
We conclude with a summary of our achievements and discuss future research questions that we would like to see addressed in Sec.~\ref{sec:conclusion}.
To make this work as helpful to the community as possible,
the arXiv version of this paper includes a simple code to generate our $\Cbp$ function.

\section{Non-perturbative broadening kernel}
\label{sec:nonpert_Cbp}
The way to UV-complete $\Cbp$, starting from an expression of a dimensionally reduced theory, was shown by \cite{Ghiglieri:2018ltw} for $\Cbp$ from $\mathcal{N}{=}4$ super-Yang-Mills theory. 
Following this example, we use the transverse momentum collision kernel $\CEQCDbp$ from lattice EQCD, measured at four different temperatures in \cite{Moore:2019lgw}, and supply it with the correct short-distance behavior in order to promote it to $\CQCDbp$, the transverse collision kernel of the full theory. Subsequently, we provide analytical expressions for the asymptotic behavior in the large- and small-$\bbp$ limits, which is then used to construct an interpolation curve of $\CQCDbp$  that can be used to calculate radiative emission rates.

\subsection{Matching for the transverse momentum broadening kernel}
Since EQCD is a low-energy effective theory for full QCD, we know that the two of them should agree%
\footnote{We use the terms IR and UV to refer to small and large momentum scales; for coordinate-space regimes, we refer to short-distance and long-distance.}
in the infrared (IR) regime, well below the hard scale $p \ll \pi T$. However, as one goes further into the ultra-violet (UV), discrepancies should arise.  Therefore, the EQCD result $\CEQCDbp$ determined in~\cite{Moore:2020wvy} cannot agree with its full QCD counterpart $\CQCDbp$, which is our ultimate object of desire.
In order to ensure the correct short-distance behavior and to keep the fully nonperturbative long-distance information from lattice EQCD at the same time, we schematically express
\begin{equation}	\label{match_strategy1}
\CQCDbp = \left( \CQCDbp - \CEQCDbp \right) + \CEQCDbp \, .
\end{equation}
The quantity in parenthesis is then computed as part
of a matching calculation between full QCD and EQCD.
We will carry out this matching calculation in transverse
momentum space and Fourier transform the result to $\bbp$-space.
Such matching calculations are free of large IR effects
because the IR regime by definition agrees between the
two theories.  Therefore the relevant scale in the
matching calculation is $2\pi T$, where the effective
coupling is (barely) perturbative and perturbation
theory should be applicable. Also, the effects of large
statistical functions, which make IR perturbative behavior
poorly behaved in thermal QCD even at weak coupling,
do not arise in this matching calculation.
Therefore we can determine this quantity perturbatively.
For the last EQCD term, we use the nonperturbative
lattice results:
\begin{equation}
\label{eq:match2}
C_{\mathrm{QCD}}(\bbp) \approx \left( C_{\mathrm{QCD}}^{\mathrm{pert}}(\bbp) - C_{\mathrm{EQCD}}^{\mathrm{pert}}(\bbp) \right) + C_{\mathrm{EQCD}}^{\mathrm{latt}}(\bbp) \,.
\end{equation}

Calculations of the perturbative terms in Eq.~(\ref{eq:match2}) are currently available at $\OO(\gs^4)$,%
\footnote{Note that $C_{\mathrm{EQCD}}^{\mathrm{pert}}(\bbp)$ and $C_{\mathrm{QCD}}^{\mathrm{pert}}(\bbp)$ obey slightly different power counting schemes; while the expansion parameter in the former case is the (four-dimensional) strong coupling constant $\gs^2$, the latter case features the (three-dimensional) expansion parameter $\gsq$. Up to subleading corrections, these two expansion parameters are related via $\gsq \approx \gs^2 T$. The full-QCD scale-dependence enters EQCD through the perturbative matching procedure \cite{Kajantie:1997tt,Laine:2005ai} that relates the EQCD effective parameters to the full-QCD coupling $\gs$ at a renormalization scale $\Lambda_{\overline{\mathrm{MS}}} \approx 341~\mathrm{MeV}$ \cite{Bruno:2017gxd}.} 
but only in transverse momentum $\qp$ space.
Therefore we will write the expression in parenthesis
in \eqref{eq:match2} in $\qtt$ space and then perform
a Fourier transform to $\bp$ space.
The perturbative QCD contribution for $\qp \gg m_D$ is given by \cite{Arnold:2008vd}
\begin{equation}	\label{Cqp_hard}
C^{\OO(\gs^4)}_\mathrm{QCD}(\qp) = \frac{\gs^4 T^3 \CR}{\qp^4} \!\int \! \frac{\d^3 p}{(2 \pi)^3} \frac{p - p_z}{p} 
\left[ 2 \CA n_\mathrm{B}(p) \left( 1 {+} n_\mathrm{B}(p') \right) + 4 \, \nf \Tf \, n_\mathrm{F}(p) \left( 1 {-} n_\mathrm{F}(p') \right) \right] .
\end{equation}
Here $\CR$ is the Casimir of the jet constituent's representation $R$ of $SU(3)$, $\CA=3$ is the respective Casimir operator of the adjoint representation, $T_\mathrm{f}=1/2$ is the normalization of the fundamental representation of $SU(3)$, $\nf$ is the number of massless fermion flavors,
and $p$ and $p' = p + \frac{\bqp^2 + 2 \bqp \hspace{-1pt} \cdot \, \bop}{2 (p - p_z)}$ are the momenta of the medium particle before/after undergoing the scattering. While \eqref{Cqp_hard} provides the correct UV limit of $C(\qp)$ in QCD, the (unphysical) IR limit of \eqref{Cqp_hard} coincides with the (unphysical) UV limit of EQCD, both from perturbation theory \cite{CaronHuot:2008ni} and the lattice \cite{Moore:2019lgw}.
We quote the full perturbative $\CEQCDbp$ at leading order \cite{Aurenche:2002wq} and next-to-leading order \cite{CaronHuot:2008ni} in \eqref{CEQCDbp_NLO} of Appendix~\ref{sec:pert_rate}. 
Here, however, we only need the large-$q_\perp$
expansion to NLO, that is, to order $\qp^{-3}$,
since this order matches the precision of the full-QCD
calculation \cite{Ghiglieri:2018ltw}.  To this order,
the EQCD value reads
\begin{equation}	\label{Cqp_subtr}
C^\mathrm{pert}_\mathrm{EQCD}(\qp) \xrightarrow{\qp \gg \mD} C^\mathrm{pert}_\mathrm{subtr}(\qp) = \frac{\CR \gs^2 T \mDsq}{\qp^4} - \frac{\CR \CA \gs^4 T^2}{16 \, \qp^3} \, .
\end{equation}
The first term in \eqref{Cqp_subtr} cancels against the IR limit of \eqref{Cqp_hard}, avoiding double-counting degrees of freedom in marrying the soft and hard contributions to $\CQCDbp$.  The second term turns into a positive
linear term in $\Cbp$.  This removes the negative
linear small-$\bp$ behavior found in EQCD, ensuring
the positivity of the full $\Cbp$.
This feature was first pointed out by
Caron-Huot \cite{CaronHuot:2008ni}.

The lattice data $C_{\mathrm{EQCD}}^{\mathrm{latt}}$
of~\cite{Moore:2020wvy} is obtained in position $(\bbp)$ space and a direct numerical Fourier transform to momentum space is a delicate issue.  Therefore we choose to
perform the matching calculation in $\bp$ space,
meaning that we need to Fourier transform the $\qp$-space
expressions we have just presented
using \eqref{subtraction_FT}, and perform the matching entirely in position space. Schematically, rewriting \eqref{match_strategy1} into
\begin{equation}	\label{match_strategy2}
C_{\mathrm{QCD}}(\bbp) \approx \int \! \frac{\d^2 \qp}{(2\pi)^2} \left( C_{\mathrm{QCD}}^{\mathrm{pert}}(\qp) - C^{\mathrm{pert}}_{\mathrm{subtr}}(\qp) \right)  \left( 1 - e^{i\bqp \cdot \mathbf{b}_{\bot}} \right) + C_{\mathrm{EQCD}}^{\mathrm{latt}}(\bbp) 
\end{equation}
is more accurate, where the Fourier transform of the difference of \eqref{Cqp_hard} and \eqref{Cqp_subtr} requires a numerical integration but is feasible with standard tools, for instance Mathematica 12. 

\begin{figure}[htbp!] 
\centering
	 \subfigure[Short-distance limit]{\includegraphics[scale=0.55]%
	 {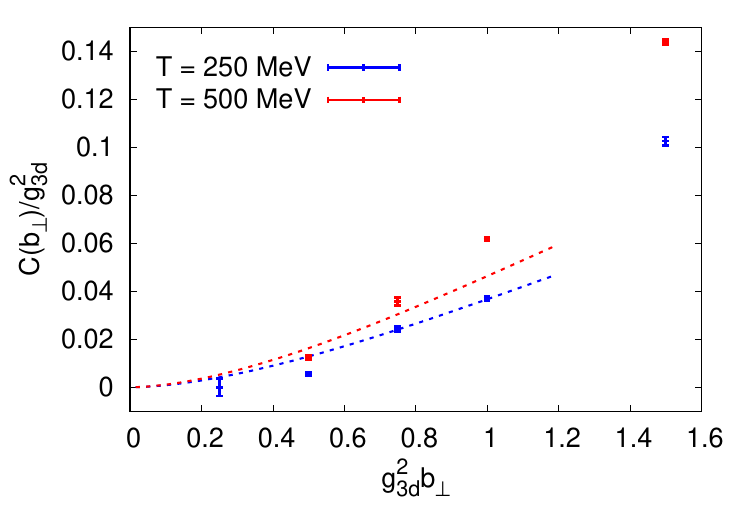}} 
	 \subfigure[Long-distance limit]{\includegraphics[scale=0.55]%
	 {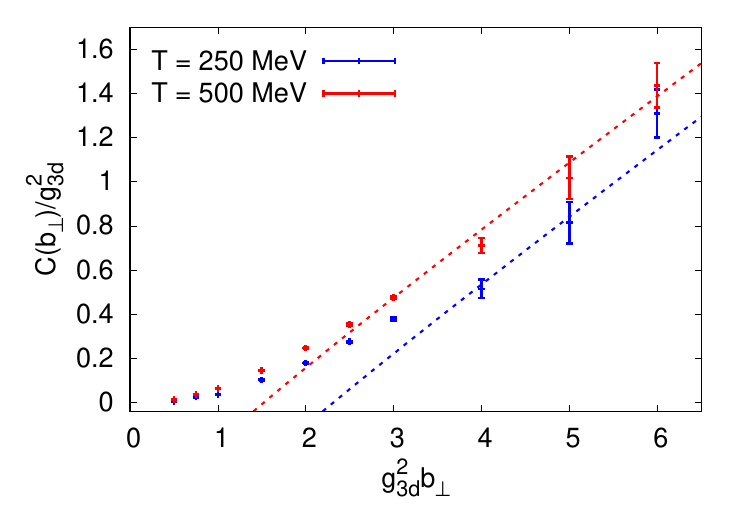}} 
	\caption{Short-distance (a) and long-distance (b) behavior of  the momentum broadening kernel $C_\mathrm{QCD}(\bbp)$. 
		Dashed lines correspond the asymptotic functional forms in \eqref{IR_limit} and \eqref{UV_limit}}
	\label{fig:asymptotics}
\end{figure}

Even though only the two lowest temperatures, $T = 250\, \mathrm{MeV}$ and $T = 500\,\mathrm{MeV}$, will be directly relevant for the subsequent computation of radiative emission rates, we provide the fully matched results for $C_{\mathrm{QCD}}(\bbp)$ in Tab.~\ref{tab:full_Cbp} for all temperatures for which $C_{\mathrm{EQCD}}^{\mathrm{latt}}(\bbp)$ was reported in~\cite{Moore:2019lgw}. Our results are cast into dimensionless ratios with the help of the three-dimensional coupling $\gsq$, and plotted in Fig.~\ref{fig:asymptotics}, where we present $\CQCDbp$ at the two temperatures of further relevance for this work together with the limiting infrared and ultraviolet behavior that we will discuss further below.

\begin{table}[ptbh] 
\centering {\small
\begin{tabular}{|c|c|c|c|c|}	
\hline
 & & & & \\[-10pt]
$\gsqb$ & $\left. \frac{\Cbp }{ \gsq} \right\vert^{N_\mathrm f = 3}_{250~\mathrm{MeV}}$ & $\left. \frac{\Cbp }{ \gsq} \right\vert^{N_\mathrm f = 3}_{500~\mathrm{MeV}}$ & $\left. \frac{\Cbp }{ \gsq} \right\vert^{N_\mathrm f = 4}_{1~\mathrm{GeV}}$ & $\left. \frac{\Cbp }{ \gsq} \right\vert^{N_\mathrm f = 5}_{100~\mathrm{GeV}}$ \\
 & & & & \\[-10pt]
\hline
$0.125$ & $-0.0011(44)$ & - & - & $-0.001(19)$ \\
$0.25$ & $0.0000(36)$ & $-0.0041(36)$ & $-0.0006(34)$ & $0.011(32)$ \\
$0.5$ & $0.00552(63)$ & $0.01244(87)$ & $0.02198(58)$ & $0.04553(49)$ \\
$0.75$ & $0.0242(11)$ & $0.0357(17)$ & $0.0559(10)$ & $0.09742(81)$ \\
$1.0$ & $0.03685(82)$ & $0.06181(61)$ & $0.09118(36)$ & $0.14850(33)$ \\
$1.5$ & $0.1025(17)$ & $0.1440(11)$ & $0.19449(94)$ & $0.28356(62)$ \\
$2.0$ & $0.1783(33)$ & $0.2467(28)$ & $0.3148(18)$ & $0.4304(12)$ \\
$2.5$ & $0.2747(41)$ & $0.3542(51)$ & $0.4443(32)$ & $0.5868(22)$ \\
$3.0$ & $0.3784(86)$ & $0.4756(46)$ & $0.5781(67)$ & $0.7425(45)$ \\
$4.0$ & $0.514(42)$ & $0.711(33)$ & $0.853(28)$ & $1.095(17)$ \\
$5.0$ & $0.815(94)$ & $1.017(96)$ & $1.09(10)$ & $1.359(13)$ \\
$6.0$ & $1.31(11)$ & $1.44(10)$ & $1.67(15)$ & $1.748(36)$ \\
\hline
$g^2$ & $3.725027$ & $2.763516$ & $2.210169$ & $1.066561$ \\
$y$ & $0.452423$ & $0.586204$ & $0.823449$ & $1.64668$ \\
\hline
$\hat{q}_{0} / \gsix$ & $0.1465(78)$ & $0.185(10)$ & $0.3136(60)$ & $0.5665(47)$ \\
\hline
\end{tabular} }
\caption{
  Results for $\frac{C_\mathrm{QCD}}{\gsq}(\bbp)$ for four 
  temperatures and a range of transverse separations $\bbp$.  
}
\label{tab:full_Cbp}
\end{table}

When comparing the lattice EQCD result $C_\mathrm{EQCD}^\mathrm{latt}(\bbp)$ in Tab.~2 of \cite{Moore:2019lgw} and the fully matched QCD result $\CQCDbp$ in our Tab.~\ref{tab:full_Cbp}, one finds that the matching only introduces marginal corrections at large $\bbp$.
However, our matching procedure in \eqref{match_strategy2} largely cures the negative dip in the short-distance regime of EQCD, where its impact is much more significant than at large distances, as expected.
All but one small-$\bbp$ values are consistent with $0$ within a single standard deviation, one value is a bit more than one standard deviation away from the positive region.
Nevertheless, not all central values of $\Cbp$ are manifestly positive.  This poses a numerical challenge later on,
and it also reflects the growing difficulty of extracting
precise data from EQCD at these short distances.
At such short distances we are instead better off trying
to handle EQCD perturbatively, so we will switch at
small $\bbp$ to the asymptotic small-separation
perturbative expression in EQCD.
Similarly, the lattice data become noisy at very
large separation, and it is best to fit them to the
expected analytical large-$\bbp$ form and to use this
analytical expression at the largest separations.
In these two limiting cases, the functional forms are available and are discussed in the following two subsections. We note that, whenever we speak of a long-distance limit in the following, we mean the limit of $\bbp \gg \xi_\mathrm{max}$, where $\xi_\mathrm{max}$ is the longest correlation length in the theory, typically $\xi_\mathrm{max} \sim \frac{1}{\gsq}$.
Similarly, whenever we speak of the short-distance limit, we mean $\bbp \ll \xi_\mathrm{min}$, typically $\xi_\mathrm{min} \sim \frac{1}{\mD}$ the smallest correlation length, such that the short-distance limit also genuinely surpasses the hard scale $\bbp < (\pi T)^{-1}$, meaning that we exceed the range of validity of EQCD and probe the matching part in this limit.

\subsection{Long-distance limit of $\Cbp$}
We first focus on the long-distance limit of $\CQCDbp$, which turns out to be the less complex issue. Beyond the known area-law form of $\CEQCDbp$, subleading asymptotic corrections are found to be important to match to the numerical data at the values of $\bbp$, where lattice calculations are feasible. Therefore, the infrared limit of the full $\Cbp$ reads
\cite{Laine:2012ht}
\begin{equation}	\label{IR_limit}
\frac{C_\mathrm{QCD}}{\gsq} (\bbp) \xrightarrow{\bbp \gg \; 1/\gsq} A + \frac{\sigma_\mathrm{EQCD}}{\gfour} \gsqb +  \frac{\gs^4 \CR}{\pi} \left[ \frac{y}{4} \left( \frac{1}{6} - \frac{1}{\pi^2} \right) + \frac{\CA}{8 \pi^2 \gs^2} \right] \log (\gsqb) \, ,
\end{equation}
where $\sigma_\mathrm{EQCD}$ is the string tension of EQCD \cite{Laine:2005ai} and $A$ is a scale-setting fitting constant. The (subleading) $\log (\dots)$-part originates from the modified Fourier-transform of the $\frac{1}{\qp^2}$-term in the small-$\qp$ expansion of \eqref{Cqp_hard}, already provided in \cite{Arnold:2008vd}. Since it is introduced by the matching procedure, this term does not show up in the IR limit of $\CEQCDbp$ \cite{Moore:2019lgw}. Meanwhile, we have re-expressed all occurrences of $\nf$ in \eqref{IR_limit} in terms of the 1-loop expression of $\mDsq = \frac{(2 \CA + \nf)}{6} g^2 T^2$ and $y = \frac{\mDsq}{\gfour} \Big|_{\bmu=\gsq}$. For $y$, in turn, we used the two-loop expression from \cite{Laine:2005ai}, giving numerical values in Tab.~\ref{tab:full_Cbp}. This procedure does not spoil the perturbative power-counting since the difference is formally subleading. In practice, we find better agreement with our data using this procedure which can be thought of as a selective resummation of higher-order contributions to the screening mass $\mDsq$. 

We plot the limiting behavior in \eqref{IR_limit} against our data for the two lowest temperatures $T = 250,\,500\,\mathrm{MeV}$ in Fig.~\ref{fig:asymptotics}(b). As one can see from Fig.~\ref{fig:asymptotics}(b), the linear contribution to \eqref{IR_limit} dominates numerically already in the displayed window of $\gsqb$. However, we find the logarithmic contribution, though numerically small, has still a non-negligible impact on our result. Just as in the case of EQCD, we see the onset of the asymptotic behavior at smaller $\gsqb$ the larger the temperature becomes, i.e.\ the smaller the (running) four-dimensional coupling $\gs$ is.

\subsection{Short-distance limit of $\Cbp$}
Evaluating $C_\mathrm{EQCD}^\mathrm{latt}(\bbp)$ at smaller $\bbp$ demands smaller lattice spacings $a$, as discretization errors of the operator occur in powers of $\frac{a}{\bbp}$~\cite{DOnofrio:2014mld}. As a consequence, the signal-to-noise ratio shrinks due to critical slowing down as one approaches $\bbp \hspace{-3pt} \to 0$. Additionally, the leading term in the small-$\bbp$-expansion of $\CEQCDbp$ is precisely an EQCD artifact and therefore cancelled by the matching, diminishing the physically relevant information further without reducing the noise. Thus, the urge for an effective analytical description is clear.

The short-distance limit of $\CQCDbp$ precisely gives rise to the momentum diffusion coefficient $\hat{q}$
\begin{equation}
\frac{C_\mathrm{QCD}}{\gsq} (\bbp) \xrightarrow{\bbp \ll \; 1/\mD} 
\frac{1}{4} \frac{\hat{q}}{\gsix}(\gsqb)^2 \, ,
\end{equation}
where $\hat{q}$ consists of a scale-dependent logarithm and constant part. Clearly, at very small $\bbp$, the logarithm is supposed to be the dominant part, and can be extracted by a modified Fourier-transform of the UV limit of \eqref{Cqp_hard} as described below. However, in keeping only the leading logarithmic terms, we do not find a smooth connection to our data, even at the smallest-$\bbp$ data points. There are two ways to resolve this issue: Either computing data points at even smaller $\bbp$ from the lattice or improving on the analytical side of the limiting behavior. Option one is certainly possible but numerically extremely costly due to the requirement of finer lattices combined with critical slowing down. Option two is somewhat easier to realize and will therefore be further pursued. 

By taking into account the constant part of the momentum diffusion coefficient, $\qhn$, which receives both perturbative and non-perturbative contributions, the short-distance limit of $\CQCDbp$ reads
\begin{equation}	\label{UV_limit}
\frac{C_\mathrm{QCD}}{\gsq} (\bbp) \xrightarrow{\bbp \ll \; 1/\mD} - \frac{\CR}{8 \pi} \frac{\zeta(3)}{\zeta(2)} \left( -\frac{1}{2 \gs^2} + \frac{3 y}{2} \right) (\gsqb)^2 \log(\gsqb) + \frac{1}{4} \frac{\qhn}{\gsix} (\gsqb)^2 \, ,
\end{equation}
where we applied the resummation of subleading contributions to the screening mass $\mDsq$ just as in the IR case.

$\qhn$ would be most straight-forwardly determined from a fit to a few of the smallest-$\bbp$-data-points of $\CQCDbp$. However, the cancellations between the EQCD result and the matching contribution make a naive fit numerically quite unstable and badly-constrained. Also, the lattice EQCD results at small $\bbp$ are strongly correlated among each other, which translates to $\CQCDbp$. 
Determining $\qhn$ directly from the lattice-EQCD-data and its correction from $\left(C_\mathrm{QCD}^{\rm pert}-C_\mathrm{subtr}^{\rm pert}\right)(\bbp)$ separately turns out to be a numerically more robust procedure, leading to the results quoted in Tab.~\ref{tab:full_Cbp}. We note that the momentum broadening coefficient of EQCD has already been determined from a small-$\bbp$-fit in \cite{Moore:2019lgw}, also taking into account the correlation between the data points. We find that, together with the matching contribution, the values of $\qhn$ in Tab.~\ref{tab:full_Cbp} are smaller than the lattice-EQCD results of~\cite{Moore:2019lgw}.

We present a comparison of our result for $C_\mathrm{QCD}(\bbp)$ in Tab.~\ref{tab:full_Cbp} to the short-distance asymptotics \eqref{UV_limit} in Fig.~\ref{fig:asymptotics}(a). Since in the accessible range of $\bbp$, deviations between the asymptotic behavior and the data points are still sizeable, the question where to switch from our data points to the asymptotic description is in fact non-trivial and introduces an uncertainty into the calculation of radiative emission rates. When parametrizing $C_\mathrm{QCD}(\bbp)$, we will therefore  keep the switching point as a parameter and finally investigate its impact on the broadening kernel.

\subsection{Numerical interpolation of lattice data}
\begin{figure}
    \includegraphics[width=\textwidth]{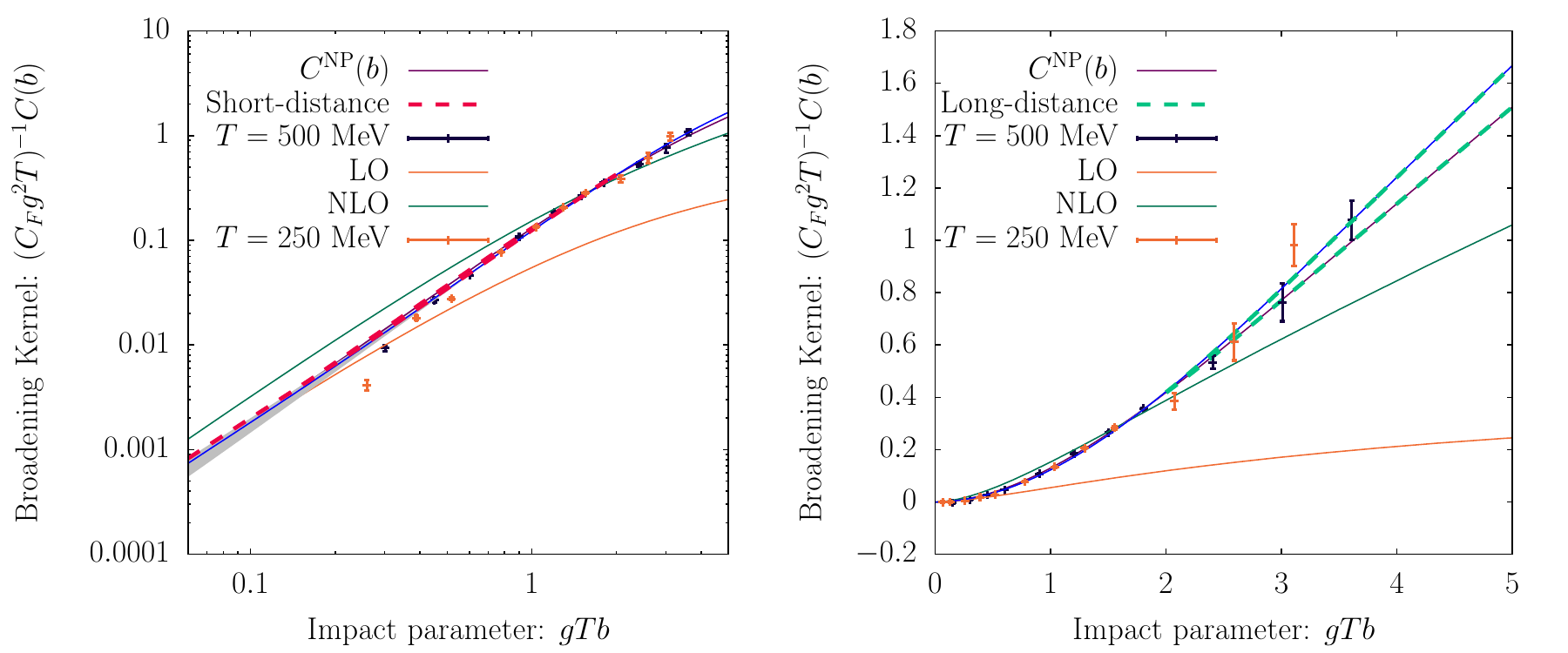}
    \caption{Non-perturbative elastic broadening kernel interpolation spline in the short-distance (left) and large-distance (right) regimes. We compare to both the short-distance limit from Eq.~(\ref{UV_limit}) and the long-distance limit from Eq.~(\ref{IR_limit}).}
    \label{fig:Spline}
\end{figure}
Next, in order to compute radiative rates, we construct a spline interpolation for the momentum broadening kernel. Guided by the limiting behaviors of $C_\mathrm{QCD} (\bbp)$, we compute several splines by varying where we switch to the asymptotic short and long distance behavior and requiring each spline to be within a standard deviation of the data points. By taking the average of the different splines we obtain the smooth spline in Fig.~\ref{fig:Spline} for the two different temperatures $T=250,500$MeV, while the gray band represents the spread of the different splines obtained. We note that the data sets for the two different temperatures show a very similar behavior when the broadening kernel $C_\mathrm{QCD} (\bbp)$ and impact parameter $\bbp$ are measured in units of $[\gs^2T]$ and $[\gs T]^{-1}$ respectively, which ultimately leads to similar radiative emission rates discussed in the next section.

Before we proceed to the application, it is also instructive to compare our results for  $C_\mathrm{QCD} (\bbp)$ with non-perturbative information from lattice EQCD with the strictly perturbative determinations of $\CQCDbp$. Specifically, at leading order (LO) $\OO(\gs^4)$, the QCD collisional broadening kernel can be expressed in momentum $(\qt)$ space as a modified version of \eqref{Cqp_hard},
\begin{align}	\label{CLO}
C^{\rm LO}_\mathrm{QCD}(\qp) = \frac{\gs^4 T^3 \CR}{\qp^2(\qp^2+m_D^2)} \int \! \frac{\d^3 p}{(2 \pi)^3} \frac{p - p_z}{p} &
\left[ 2 \CA n_\mathrm{B}(p) \left( 1 + n_\mathrm{B}(p') \right) \right.
\nonumber \\  & \left. {} + 4 \, \nf T_\mathrm{f} \, n_\mathrm{F}(p) \left( 1 - n_\mathrm{F}(p') \right) \right] \, ,
\end{align}
which at LO is valid for all $\qp$ \cite{Arnold:2008vd}. Next-to-leading order (NLO) corrections are of $\OO(\gs^5)$; they arise from infrared corrections that are suppressed by an additional factor of $\mD \sim \gs$ and can be calculated within EQCD~\cite{CaronHuot:2008ni}. Hence, to obtain the NLO result, we follow the same matching procedure as for the non-perturbatively determined lattice EQCD results in \eqref{match_strategy2} and obtain $C^{\rm NLO}_\mathrm{QCD}$ in momentum space as
\begin{equation}	\label{CNLO}
C^{\rm NLO}_\mathrm{QCD}(\qp)= \left( C_{\mathrm{QCD}}^{\mathrm{pert}}(\qp) - C^{\mathrm{pert}}_{\mathrm{subtr}}(\qp) \right)  + C_{\mathrm{EQCD}}^{\mathrm{LO}}(\qp) + C_{\mathrm{EQCD}}^{\mathrm{NLO}}(\qp) 
\end{equation}
which really corresponds to supplementing the $\OO(\gs^4)$-result \eqref{Cqp_hard} with the appropriate infrared contributions $C_{\mathrm{EQCD}}^{\mathrm{LO}}(\qp) + C_{\mathrm{EQCD}}^{\mathrm{NLO}}(\qp) - C^{\mathrm{pert}}_{\mathrm{subtr}}(\qp)$ provided in \eqref{CEQCDbp_NLO} of Appendix \ref{sec:pert_rate}. By comparing the different results for the broadening kernel $C_\mathrm{QCD} (\bbp)$ in Fig.~\ref{fig:Spline}, one observes that the leading order result in (\ref{CLO}) provides a reasonable description of the extrapolated spline for small values of $\gs T \bbp$; the next-to-leading order (NLO) result features a significantly larger value of $\hat{q}$ but has the same qualitative infrared behavior as the non-perturbatively determined $C_\mathrm{QCD} (\bbp)$.


\section{Medium induced splitting rates} 
\label{sec:inelastic_rates}

We will now illustrate how the results for the collisional broadening kernel discussed in the previous section can be employed to compute radiative emission rates for highly energetic particles.
While different formalisms exist to compute medium induced radiation in QCD matter\footnote{See \cite{Armesto:2011ht} for a comparison of the different approaches.} \cite{Baier:1996kr,Zakharov:1996fv,Gyulassy:2000er,Arnold:2001ms},  we emphasize that any of these approaches can make use of the non-perturbative elastic scattering rate. We choose to follow the formalism of Arnold, Moore and Yaffe (AMY) \cite{Arnold:2001ms}, which can be formulated directly in coordinate space and provides an effective rate $\d \Gamma_{ij}/ \d z(P,z)$, which corresponds to the rate at which particle $i$ with energy $P$ radiates particle $j$ with energy $\omega = zP$ in an infinite medium. In addition to radiative emission rates determined from the non-perturbative momentum broadening kernel $C_{\rm QCD}(\bbp)$, we will also consider the rates obtained with leading $C_{\rm QCD}^{\rm LO}(\bbp)$ and next-to-leading order $C_{\rm QCD}^{\rm NLO}(\bbp)$ results for momentum broadening.

\subsection{Effective splitting rate }
The starting point for the calculation of the inelastic splitting rates  $\d \Gamma_{ij}/ \d z(P,z)$ is the following expression
\begin{eqnarray}
\frac{\d \Gamma_{ij}}{\d z} (P,z) = \frac{\alpha_s P_{ij}(z)}{[2Pz(1{-}z)]^2} \int \! \frac{\d ^2 \ptt}{(2\pi)^2} ~\text{Re} \left[ 2\ptt \cdot \mathbf{g}_{(z,P)}(\ptt) \right]\;,
\end{eqnarray}
where we follow the notation of P.~Arnold in Appendix A of \cite{Arnold:2008iy}. Here $P_{ab}(z)$ are the Dokshitzer-Gribov-Lipatov-Altarelli-Parisi (DGLAP) splitting functions
\begin{eqnarray}
    P_{gg}(z) &=& 2 \CA \frac{[1-z(1{-}z)]^2}{z(1{-}z)}\;,\quad
    P_{qg}(z)  = \CF \frac{1+(1{-}z)^2}{z}\;,\quad
    P_{gq}(z) = \frac{1}{2} \left(z^2+(1{-}z)^2\right) \;. \nonumber\\
\end{eqnarray}
The function $\mathbf{g}_{(z,P)}(\ptt)$, which encodes the current-current correlator, satisfies the following integral equation:
\begin{eqnarray}
    \label{eq:AMY}
    2\ptt &=& i \delta E(z,P,\ptt) \mathbf{g}_{(z,P)}(\ptt) + \int \! \frac{\d ^2\qtt}{(2\pi)^2}~\bar{C}(q_\perp)   \\
    && \hspace{120pt} \times \left\{ C_{1} \left[ \mathbf{g}_{(z,P)}(\ptt) - \mathbf{g}_{(z,P)}(\ptt -\qtt) \right] \right. \nonumber \\
    && \hspace{140pt} \left. + \, C_{z} \left[ \mathbf{g}_{(z,P)}(\ptt) - \mathbf{g}_{(z,P)}(\ptt -z\qtt) \right] \right. \nonumber \\
    && \hspace{140pt} \left. + \, C_{1-z} \left[ \mathbf{g}_{(z,P)}(\ptt) - \mathbf{g}_{(z,P)}(\ptt -(1{-}z)\qtt) \right] \right\} \, . \nonumber
\end{eqnarray}
The energy difference $\delta E(z,P,\ptt)$ is written
\begin{eqnarray}
\delta E(z,P,\ptt) = \frac{\ptt^2}{2Pz(1{-}z)} +M_{\rm eff}(z,P)\;,
\end{eqnarray}
where $M_{\rm eff}(z,P)$ is given in terms of the asymptotic masses $m^2_{\infty,(1,z,1-z)}$ of the particles with momentum fractions $1,z,1-z$ as
\begin{eqnarray}
M_{\rm eff}(z,P)=  \frac{m^2_{\infty,(z)}}{2zP}  + \frac{m^2_{\infty,(1{-}z)}}{2(1{-}z)P} -\frac{m^2_{\infty,(1)}}{2P}.
\end{eqnarray}
For the asymptotic masses we use the leading order results given by
\begin{eqnarray}
    m^2_{\infty,g}=\frac{\mDsq}{2}= \frac{\gs^2T^2}{6}\left( \CA +\frac{\nf}{2}\right)\;,\qquad
    m^2_{\infty,q}= \CF \frac{\gs^2 T^2}{4}\;.
\end{eqnarray} 
The color factors are given by
\begin{eqnarray}
\label{colorfactors}
C_{1}=\frac{1}{2} \left( C^{R}_{z} + C^{R}_{1-z} - C^{R}_{1} \right)\;, \nonumber &\quad&
C_{z}=\frac{1}{2} \left( C^{R}_{1-z} + C^{R}_{1} - C^{R}_{z} \right)\;,  \nonumber \\
C_{1-z}&=&\frac{1}{2} \left( C^{R}_{1} + C^{R}_{z} - C^{R}_{1-z} \right)\;, 
\end{eqnarray}
where $C^{R}_{(1,z,1-z)}$ denote the Casimir of the representation of the particle carrying momentum fraction $1,z,1-z$, i.e. $C^{R}=\CF$ for quarks and $C^{R}=\CA$ for gluons. Since the color factors have been factored out, the rate $\bar{C}(q)$ in Eq.~(\ref{eq:AMY}) denotes the elastic scattering rate stripped of its color factor.%
\footnote{%
Technically, we make an approximation when we use $\bar{C}(q_\perp)$
and \eqref{colorfactors} in \eqref{eq:AMY}.  Our expression
for $\bar{C}(q_\perp)$ is for a 
fundamental-representation particle picking up
transverse momentum $q_\perp$ in the medium, and is determined
by the Wilson line we described;
but to compute the evolution of a mixed state
with an emitter in one amplitude and an emitter and a gluon
in the conjugate amplitude, we really need a 3-Wilson-line
object, as described in Ref.~\cite{CaronHuot:2008ni}.
This factorizes into a combination of Wilson-line-pair
contributions as shown in \eqref{colorfactors} up to corrections
which are at most NNLO.  Theoretically, this three-Wilson-line
object could also be directly computed in EQCD, but it
would be more challenging, and in particular we would
need a separate numerical calculation for each pair $(b_\perp,x)$.
Here we are assuming that the NNLO corrections are small
and the 3-Wilson-line object factorizes as shown.
As far as we know, \textsl{every} treatment in the literature
makes this same approximation!
}

We solve Eq.~(\ref{eq:AMY}) in impact-parameter space for the splitting rates using the numerical procedure outlined in the Appendix~\ref{sec:numerical} to obtain the rates of medium induced $g \to gg$, $q \to qg$ and $g \to q\bar{q}$ splittings and supply the software as part of the arXiv submission package. Since the fully resummed AMY rate includes both the high-energy limit where the Landau-Pomeranchuk-Migdal (LPM) effect is prominent, and the low-energy limit which follows a Bethe-Heitler (BH) rate, we will briefly discuss these limits before addressing the numerical results in more detail.

\subsection{Bethe-Heitler regime}
 When the typical momentum of splitting is small $(Pz(1{-}z) \ll \omega_{\rm BH} \sim T )$, the formation time of the radiation is small and interference between scatterings can be neglected. In this so-called Bethe-Heitler regime, one can then solve the rate Eq.~(\ref{eq:AMY}) in an opacity expansion, corresponding to expansion in the number of elastic scatterings with the medium. By consider the limit of a single scattering, we obtain the following semi-analytic expressions for the rates (c.f. Appendix \ref{sec:BH})
 \begin{align}
    \label{eq:BHRate}
    \frac{\d \Gamma_{g\to gg}^{BH}}{\d z} (P,z) = & 
    \gs^4  T P_{gg}(z) \times \\ 
    & \left[  \frac{C_A}{2} Q\left( \mut^{2}_{g\to gg}(z) \right)  + \frac{\CA}{2} Q\left( \mut^{2}_{g\to gg}(z)/z^2 \right) + \frac{\CA}{2} Q\left( \mut^{2}_{g\to gg}(z)/\bar{z}^2 \right) \right], \nonumber \\ 
    \frac{\d \Gamma_{q\to gq}^{BH}}{dz} (P,z) = & 
    \gs^4 T P_{qg}(z) \times \nonumber \\
    & \left[\frac{\CA}{2} Q\left( \mut^{2}_{q\to gq}(z) \right)  +(\CF-\frac{\CA}{2}) Q\left( \mut^{2}_{q\to gq}(z)/z^2 \right) +\frac{\CA}{2} Q\left( \mut^{2}_{q\to gq}(z)/\bar{z}^2 \right) \right], \nonumber \\ 
    \frac{\d \Gamma_{g\to qq}^{BH}}{\d z} (P,z) = &
     \gs^4 T  P_{gq}(z) \times \nonumber \\
     & \left[ (\CF - \frac{\CA}{2}) Q\left( \mut^{2}_{g\to qq}(z) \right) + \frac{\CA}{2} Q\left( \mut^{2}_{g\to gq}(z)/z^2 \right) +\frac{\CA}{2} Q\left( \mut^{2}_{q\to gq}(z)/\bar{z}^2 \right) \right],\nonumber
\end{align}
where, denoting $a=m^2_{\infty,q}/m_D^2$, one has
\begin{align}
\mut^{2}_{g\to gg}(z)&=\frac{1-z(1{-}z)}{2}\;,  & \mut^{2}_{q\to gq}(z)&=\frac{z}{2}+a(1{-}z)^2\;, & \mut^{2}_{g\to qq}(z)& =\frac{2a-z(1{-}z)}{2}\;,
\end{align}
and
\begin{eqnarray}
\label{eq:QBHDef}
Q(\mut^2)=  \frac{\mDsq}{2\pi \gs^2 T} \int \! \frac{\d ^2 \ptt}{(2\pi)^2} \int \! \frac{\d ^2\qtt}{(2\pi)^2} ~ \bar{C}(m_D\qtt)  \left[ \frac{\ptt}{\ptt^2+\mut^2} - \frac{(\ptt-\qtt) }{(\ptt-\qtt)^2+\mut^2}  \right]^2\;.
\end{eqnarray}
While the above relation is formulated in momentum space, the  integral defining $Q(\mut^2)$ in Eq.~(\ref{eq:QBHDef}) can also be evaluated using the kernel in position space as show in Appendix \ref{sec:BH}.

\subsection{Deep LPM regime}
Conversely, in the limit of a very high-energy parton $(P\gg T)$ traversing a thick medium, the typical number of rescatterings within the formation time of bremsstrahlung can be large, indicating that interferences between many soft scatterings which contribute to the total transverse momentum transfer during the formation of the radiation need to be considered. Simplifications occur in the limit $Pz(1{-}z) \gg \omega_{\rm BH} \sim T$, where the splitting probes the small $\bbp$ behavior of the momentum broadening kernel which can be expressed as
\begin{eqnarray}
C(\bbp) = -\frac{\gs^4 T^3}{16\pi}\mathcal{N}\bbp^2 \log( \xi \mDsq \bbp^2/4)
\end{eqnarray}
where $\mathcal{N} = \frac{\zeta (3)}{\zeta(2)} \left( 1 + \tfrac{\nf}{4} \right)$. In accordance with the discussion in Sec.~\ref{sec:nonpert_Cbp}, the coefficient $\frac{\gs^4T^2}{16\pi}\mathcal{N} \bbp^2$ gives the leading logarithmic behavior $\bbp \log(\bbp^2)$ and the coefficient $\xi$ captures the $\bbp^2$ behavior. Specifically at for the LO kernel $\xi_{\rm LO}=e^{2\gamma_\mathrm{E}-2}\simeq 0.429313$ can be determined analytically, while for the NLO and non-perturbative kernels, we obtain $\xi_{\rm NLO} \simeq 1.355 \cdot10^{-3}$ and $\xi_{\rm NP} = 4\frac{\gs^4 T^2 }{\mDsq} e^{-4\pi\frac{  \hat{q}_0}{\gs^4 T^3 \mathcal{N}}}\simeq 0.1702$ from a fit of the small $\bbp$ behavior. Following \cite{Arnold:2008zu}, the rate equation can be solved iteratively in an inverse logarithmic expansion to obtain\footnotemark
\footnotetext{ The rate in our notation is related to the notation in \cite{Arnold:2008zu} as $\frac{\d \Gamma_{a\to bc}}{\d z}(P) = \frac{(2\pi)^3}{P\nu_a}\gamma^{a}_{bc}(P|zP,\bar{z}P)$ with $\nu_g = 2(\Nc^2-1)$ and $\nu_q = 2\Nc$.}
\begin{eqnarray}
    \label{eq:LPMRate}
    \frac{\d \Gamma_{a\to bc}}{\d z}(P)&=&\frac{\gs^2}{16\pi^2\sqrt{2} Pz(1-z)}~P_{ab}(z)~\mDsq \mu^2_\perp(P,z)\;,
\end{eqnarray}
where $\mu^2_\perp(P,z)$ is self-consistently determined from 
\begin{eqnarray}
    \mu^2_\perp(P,z) &=& \frac{\gs^2 T^2\mathcal{N}}{\mDsq}\frac{\gs T}{\mD} \left( \frac{2}{\pi} z(1{-}z)\frac{P}{T} \right)^{1/2}
    \bigg( C_1 \log\left( \frac{\alpha  \mu_\perp^2}{\xi}\right)  \nonumber \\
    &&\qquad \left. + \, C_z z^2 \log\left(\frac{\alpha  \mu_\perp^2}{\xi z^2}\right) + \, C_{1-z} (1-z)^2 \log\left(\frac{\alpha  \mu_\perp^2}{\xi(1{-}z)^2}\right)\right)^{1/2}\;,
\end{eqnarray}
with  $\alpha= e^{\gamma_e+\pi/4}$.

\subsection{Results}

\begin{figure}[h!]
    \includegraphics[width=0.98\textwidth]{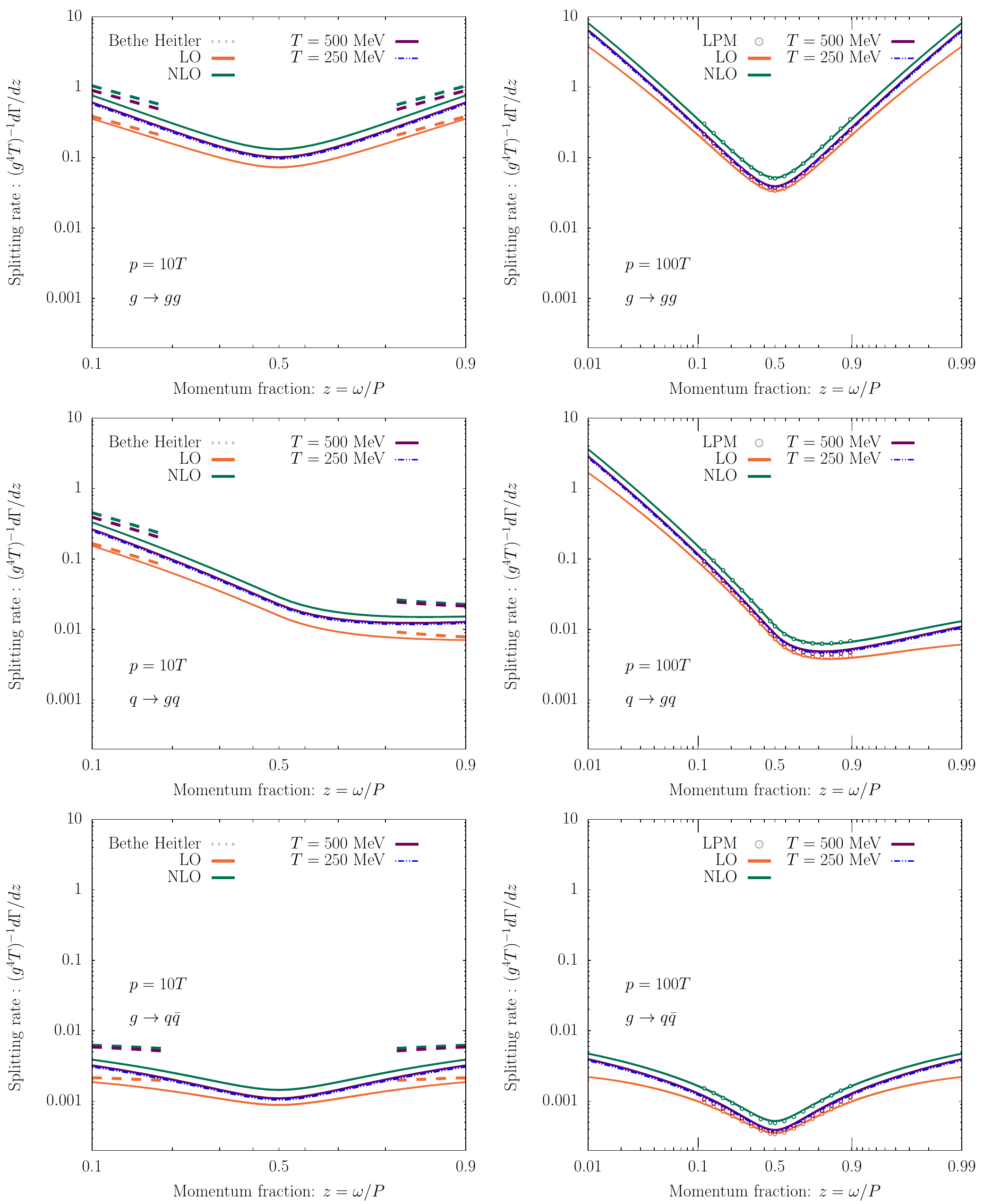}
    \caption{Splitting rate at $T=250$MeV (dashed blue lines) and $T=500$MeV (full purple lines) for the processes $g \rightarrow gg$ (top),  $q \rightarrow gq$ (middle), $g \rightarrow qq$ (bottom). Different columns correspond to parent energies $p=10T$ (left) and $p=100T$ (right). We compare with rates computed using the perturbative leading order (orange) and next-to-leading order (green) elastic broadening kernels. The Bethe Heitler rates and LPM rates are shown with dashed lines and circles respectively, using the color of the corresponding kernel. }
    \label{fig:SplittingRates500}
\end{figure}

\begin{figure}[h!]
    \includegraphics[width=0.98\textwidth]{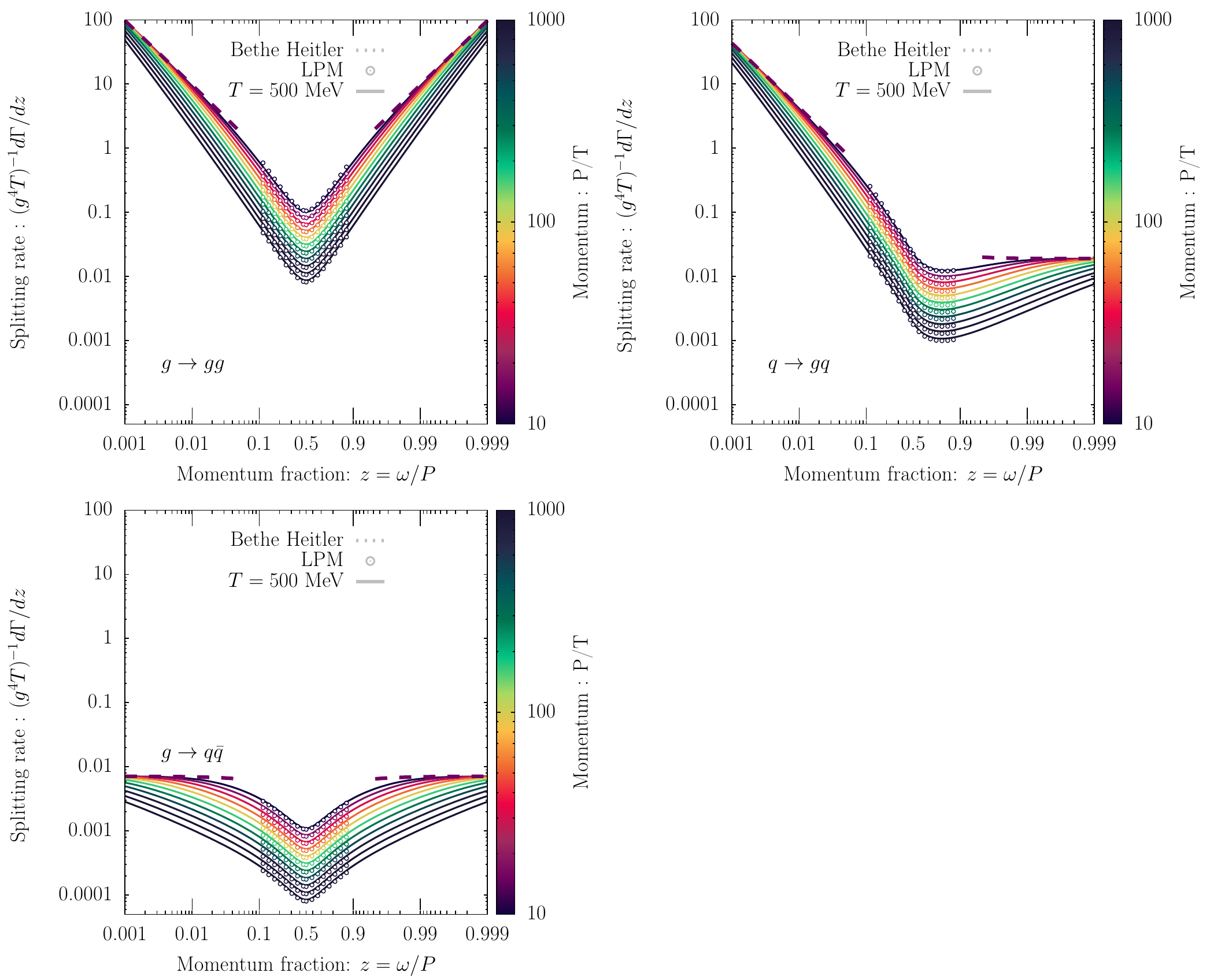}
    \caption{Momentum dependence of the splitting rate at $T=500$MeV (dashed blue lines) for the processes $g \rightarrow gg$ (top),  $q \rightarrow gq$ (right), $g \rightarrow qq$ (bottom). Dashed lines and open circles correspond to the approximate rates in Bethe Heitler regime (\ref{eq:BHRate}) and the deep LPM regime (\ref{eq:LPMRate}). }
    \label{fig:MomentumRate}
\end{figure}

Numerical results for the medium-induced splitting rates are presented in Fig.~\ref{fig:SplittingRates500}, where we show the rates for the non-perturbative broadening kernel $C_{\rm QCD}(\bbp)$ at $T=250,500$MeV (left/right columns), along with the corresponding results obtained for the leading order ($C_{\rm QCD}^{\rm LO}(\bbp)$) and next-to-leading order ($C_{\rm QCD}^{\rm NLO}(\bbp)$) determinations of the collisional broadening kernel.
Different panels in Fig.~\ref{fig:SplittingRates500} show the rates for the $g\to gg$ (top),$q \to qg$ (center) and $g \to q\bar{q}$ (bottom) processes, for different parton energies $p=10T$ (left) and $p=100T$ (right).  The momentum dependence of the rate is shown in Fig.~\ref{fig:MomentumRate}, for the non-perturbative kernel at $T=500$MeV, using the color palette in a logarithmic scale to distinguish between different momentum of the parent particle $p=10-1000T$. In both figures, we also show the Bethe-Heitler rates in Eq.~(\ref{eq:BHRate}) (dashed lines) and the deep LPM rates (circles) Eq.~(\ref{eq:LPMRate}).

Starting with the rates in Fig.~\ref{fig:SplittingRates500}, one observes that the non-perturbatively determined splitting rates for the two different temperatures do not display any remarkable difference, leading to basically the same emission rates in units of $[g^4T]$. As expected, the momentum dependence of the rate clearly displays LPM suppression at large typical momentum $Pz(1{-}z) \gg T$ as well as an unsuppressed Bethe-Heitler rate in the other limit $Pz(1{-}z) \ll T$ seen in Fig.~\ref{fig:MomentumRate}.
We also observe that at low energy $z(1{-}z)E \ll T$ where the large impact parameter (small momentum transfer) is more important, the non-perturbative result is closer to the NLO rate as they both have a similar behavior at large impact parameter. 
Despite the apparent non-convergence of the perturbative series for $\Cbp$ observed in \cite{CaronHuot:2008ni}, it is interesting to observe that the splitting rates for the non-pertubatively determined $\Cbp$ mostly fall between the LO and NLO results and it would be interesting to compare them to higher order perturbative calculations in the future.
Conversely, in the LPM suppressed regime at high energy  $z(1{-}z)E \gg T$, where small impact parameter (large momentum transfer) is relevant, the non-perturbative rate is closer to the LO rate which again agrees with the behavior of the elastic kernel (cf. 
Fig.~\ref{fig:Spline}).

\section{Conclusion}
\label{sec:conclusion}

Despite small values of the strong coupling $\alphas$ at large temperature, perturbative calculations of transport phenomena in
the QGP can receive large non-perturbative contributions due to the famous infrared problem of finite temperature QCD. In the present work, we investigated the impact of non-perturbative contributions on jet-medium interactions, by incorporating non-perturbative contributions to the collisional broadening kernel $C(\bbp)$, which determines the rate of medium induced splittings. We appended the non-perturbative data from lattice EQCD that dominates in the large- and intermediate-$\bbp$-regime with the
correct small-$\bbp$-limit via a matching calculation, where we subtracted
the perturbative small-$\bbp$ limit of $\CEQCDbp$ and replaced it with the perturbative $\CQCDbp$ of full QCD in Sec.~\ref{sec:nonpert_Cbp}. 
Beyond the range of $\bbp$ in which lattice data is available, we provide analytical expressions for the short-distance and long-distance limits of $\CQCDbp$ and reconstruct $\Cbp$ as a function of $\bbp$ over the entire range of values, by interpolating between our data points with a sufficiently 
smooth spline that recovers the two limiting cases. We find that for $T=250{\rm MeV}$ and $500{\rm MeV}$ the leading temperature dependence of $\Cbp$ can be scaled out, such that $C(g T \bbp)/(g^2 T \bbp)$ is approximately independent of the temperature, in the relevant regime explored in this study. In order to facilitate the use of $C_{\rm QCD}(\bbp)$ in phenomenological studies of jet quenching, we provide C/C++ routines of the interpolating spline as part of the arXiv submission.

Subsequently, in Sec.~\ref{sec:inelastic_rates}, we calculated the medium-induced splitting 
rate based on our non-perturbatively determined $\Cbp$. For the sake of simplicity, we 
restrict ourselves to the simpler case of medium induced splittings in an infinite medium, although we see no conceptual problem 
in generalizing our approach to a medium of finite extent. 
We further compared our non-perturbative splitting rates to a number of other common approximations in the field: the full rates using leading and next-to-leading order $\Cbp$, the simplified rates in the deep LPM regime, 
and the Bethe-Heitler approximation. We find substantial deviations from all these cases in the physically relevant 
ranges of energies and momentum fractions even though our rates reproduce all of the mentioned approximations deeper in the 
respective limit. Our results make a compelling case for incorporating non-perturbative large-$\bbp$ physics into the 
computation of medium-induced splitting rates and jet observables in the long run.
In the near future, a generalization of our approach to more realistic finite-medium considerations seems natural. 
Furthermore, non-perturbative effects in the \textsl{longitudinal} momentum diffusion were recently calculated \cite{Moore:2020wvy} and -- after a similar matching 
procedure as outlined for $\Cbp$ above -- await application in a subsequent, improved non-perturbative calculation of 
medium induced splittings rates. 

With this in hand, a computation of experimentally measurable quantities like $R_\mathrm{AA}$ would be in reach,
allowing to see if non-perturbative jet-medium-interactions can indeed explain the suppression of 
large-transverse-momentum jets in heavy-ion collisions compared to proton-proton collisions.

\section*{Acknowledgements}
We would like to express our gratitude to Jacopo Ghiglieri for fruitful discussions in the early stages of this problem,
for careful comments on the original draft,
and for providing us with a Mathematica notebook performing the numerical Fourier transform of \eqref{Cqp_hard}.
We thank Shuzhe Shi for comments on the draft.
This work is supported by the Deutsche Forschungsgemeinschaft (DFG, German Research Foundation) through the CRC-TR 211 ’Strong-interaction matter under extreme conditions’– project number 315477589 – TRR 211. N.\ S.\ acknowledges support from Academy of Finland grants 267286 and 320123.

\appendix

\section{Perturbative results for collisional broadening}
\label{sec:pert_rate}
We define here the leading (LO) and next-to-leading (NLO) order broadening kernels. Following \cite{Arnold:2008vd}, the LO is given by the following integral:
\begin{eqnarray}
    C^{\rm LO}_{\rm QCD}(\qp) &=& \frac{ \gs^4 \CR}{\qt^2(\qt^2+ \mDsq)} \int \! \frac{\d ^3p}{(2\pi)^3}
    \frac{p-p_z}{p}
    \left[ 2 \CA n_\mathrm{B}(p)(1 + n_\mathrm{B}(p')) \right.\nonumber \\ 
    && \hspace{130pt} \left.+ 4 \nf \Tf n_\mathrm{F}(p)(1 - n_\mathrm{F}(p'))\right]\;,
\end{eqnarray}
with $p' = p + \frac{\bqp^2 + 2 \bqp \hspace{-1pt} \cdot \, \bop}{2 (p - p_z)}$. The kernel displays the following asymptotic behaviors:
\begin{eqnarray}
C^{\rm LO}_{\rm QCD}(\qp) &=& \gs^2 T \CR
    \begin{cases}
        \frac{ \mDsq - \gs^2 T^2 \CA \frac{\qp}{16 T}}{\qp^2 (\qp^2 + \mDsq)} \;, & \qp \ll \gs T \;,\\\\
        \frac{ \gs^2 T^2}{\qp^4} \frac{\zeta (3)}{\zeta(2)} \left( 1 + \tfrac{\nf}{4} \right)\;, & \qp \gg \gs T \;.\\
    \end{cases}
\end{eqnarray}
Similarly to the treatment of the non-perturbative kernel, the NLO broadening kernel is computed using perturbative results for the soft contributions from EQCD and supplying the hard contribution by the matching \eqref{Cqp_hard} \cite{CaronHuot:2008ni}. Specifically,
\begin{eqnarray}
    C^{\rm NLO}_{\rm QCD}(\qp) &=&C^{\rm LO }_{\rm EQCD}(\qp)+C^{\rm NLO }_{\rm EQCD}(\qp) + C_{\rm QCD}^{\rm pert}(\qp) - C_{\rm subtr}^{\rm pert}(\qp)\;,
\end{eqnarray}
where the leading and next-to-leading order contributions from soft modes are given by
\begin{eqnarray}	\label{CEQCDbp_NLO}
    C^{\rm LO }_{\rm EQCD}(\qp) &=& \CR \gs^2 T \frac{\mDsq}{\qt^2(\qt^2 + \mDsq)}\;,\\
    \frac{C^{\rm NLO }_{\rm EQCD}(\qp)}{\gs^4T^2 \CR \CA} &=&
    \frac{7}{32 \qt^3} +
    \frac{  {-} \mD -2\frac{\qt^2{-} \mDsq}{\qt}
    \tan^{-1} \left(\frac{\qt}{ \mD}\right) }
    {4\pi(\qt^2{+} \mDsq)^2}
    +\frac{ \mD - \frac{\qt^2{+}4 \mDsq}{2\qt}
    \tan^{-1}\left(\frac{\qt}{2 \mD}\right)}
    {8\pi \qt^4}
    \nonumber\\
    &&
    - \frac{\tan^{-1} \left(\frac{\qt}{ \mD}\right) }
    {2\pi \qt(\qt^2 + \mDsq)}
    +\frac{\tan^{-1} \left(\frac{\qt}{2 \mD}\right) }{2\pi \qt^3}
    \nonumber\\
    && +
    \frac{ m_D}{4\pi(\qt^2{+} \mDsq)}\left[
    \frac{3}{\qt^2{+}4 \mDsq}
    -\frac{2}{(\qt^2{+} \mDsq)}
    -\frac{1}{\qt^2} \right]\;.
\end{eqnarray}
and $C_{\rm subtr}^{\rm pert}(\qp)$ from \eqref{Cqp_subtr} cancels the IR divergence of the hard contribution and the UV behavior of the soft NLO contribution.

\section{Numerical implementation}
\label{sec:numerical}
Several strategies have been developed in the literature to obtain a numerical solution for the calculation of the inelastic splitting rates. We follow the strategy of \cite{Anisimov:2010gy}, and provide a detailed outline of the procedure below.
\subsection{Impact parameter space}

In order to solve for the splitting function in Eq.~(\ref{eq:AMY}), we switch from transverse momentum  $\ptt$ to transverse impact parameter $\bp$ space according to
\begin{eqnarray}
\mathbf{f}_{(z,P)}(\bp) = \frac{1}{16 P^2z^2(1{-}z)^2} \int \! \frac{\d ^2\ptt}{(2\pi)^2}~e^{i \bp \ptt}\mathbf{g}_{(z,P)}(\ptt)
\end{eqnarray}
such that
\begin{eqnarray}
\mathbf{g}_{(z,P)}(\ptt) = 16 P^2z^2(1{-}z)^2~\int \! \d ^2\bp ~e^{-i \bp \ptt}~\mathbf{f}_{(z,P)}(\bp) 
\end{eqnarray}
where for later convenience we have absorbed a pre-factor $16 P^2 z^2 (1{-}z)^2$ into the definition, we need to calculate
\begin{eqnarray}
\frac{\d \Gamma_{ij}}{\d z} (P,z) =\alpha_s P_{ij}(z)\text{Im} \left[ 8\nabla_{\bp} \cdot \mathbf{f}_{(z,P)}(\bp) \right]
\end{eqnarray}
we get $\ptt \to -i \nabla_{\bp}$ and $\ptt^2 \to -  \nabla_{\bp}^2$ such that
\begin{eqnarray}
\frac{-2i}{16 P^2z^2(1{-}z)^2}~\nabla_{\bp} \delta^{(2)}(\bp) &=& i\left( M_{\rm eff}(z,P)- \frac{\nabla_{\bp}^2}{2 P z(1{-}z)} \right)~\mathbf{f}_{(z,P)}(\bp) \nonumber \\
&&+ \left[ C_{1} \bar{C}(\bp) + C_{z} \bar{C}(z\bp) + C_{1-z} \bar{C}((1{-}z)\bp) \right]~\mathbf{f}_{(z,P)}(\bp)\;, \nonumber \\\label{eq:ODE} 
\end{eqnarray}
where $\bar{C}(\btt)$ denotes the elastic scattering rate in impact parameter space stripped of its color factor.

\subsection{General strategy}
Since for $\bp \to 0$ the contributions from the collision term vanish, the structure of the solution near the origin is already contained in the "free" solution $\mathbf{f}^{(0)}_{(z,P)}(\bp)$. By matching the most singular terms near the origin one has
\begin{eqnarray}
\frac{-2i}{16 P^2z^2(1{-}z)^2} \nabla_{\bp} \delta^{(2)}(\bp) &=&- i\frac{\nabla_{\bp}^2}{2 P z(1{-}z)}~\mathbf{f}^{(0)}_{(z,P)}(\bp)\;,
\end{eqnarray}
which yields
\begin{eqnarray}
\lim_{\bp \to 0} \mathbf{f}^{(0)}_{(z,P)}(\bp) =  \frac{1}{8\pi P z(1{-}z)} \frac{\bp}{|\bp|^2} 
\end{eqnarray}
as can be easily seen from considering the two dimensional version of Gauss law. Denoting $\bbp=|\bp|$ and expressing 
\begin{eqnarray}
\mathbf{f}^{(0)}_{(z,P)}(\bp) = \bp f^{(0)}_{(z,P)}(\bbp)\;,
\end{eqnarray}
such that component wise
\begin{eqnarray}
\nabla_{\bp}^{i} \nabla_{\bp}^{i}  \bp^{j} f^{(0)}_{(z,P)}(\bbp) = \bp^{j} \left( f^{(0)''}_{(z,P)}(\bbp)+ \frac{3}{\bbp} f^{(0)'}_{(z,P)}(\bbp) \right)
\end{eqnarray}
we then have to solve
\begin{eqnarray}
\left[\partial_{\bbp}^{2} + \frac{3}{\bbp} \partial_{\bbp} - 2P z (1{-}z) M_{\rm eff}(z,P) \right] f^{(0)}_{(z,P)}(\bbp) = 0\;,
\end{eqnarray}
for $\bbp>0$ with the boundary conditions
\begin{eqnarray}
\lim_{\bbp \to 0}  {f}^{(0)}_{(z,P)}(b) = \frac{1}{8\pi P z (1{-}z) \bbp^2} \;, \qquad  \lim_{\bbp \to \infty}  {f}^{(0)}_{(z,P)}(\bbp)=0\;,
\end{eqnarray}
 to determine the "free'' solution. By implementing the correct boundary conditions, one finds that the appropriate solution to Bessel's equation is given by
 \begin{equation}
 {f}^{(0)}_{(z,P)}(\bbp)=  \frac{1}{8\pi P z (1{-}z) \bbp^2} ~ \sqrt{2 P z(1{-}z) M(z,P)}\bbp~K_{1}\Big(\sqrt{2 Pz(1{-}z)M(z,P)} \bbp \Big)\;.
 \end{equation}
 where $K_{\nu}(z)$ denotes the modified Bessel function of rank $\nu$ and argument $z$. Note that the free solution $ {f}^{(0)}_{(z,P)}(\bbp)$ is entirely real and hence does not contribute to the expression for the radiation rate.

Based on the "free'' solution ${f}^{(0)}_{(z,P)}(\bbp)$, the solution $f_{(z,P)}(\bbp)$ to the full evolution equation can be conveniently expressed as
 \begin{eqnarray}
 f_{(z,P)}(\bbp)= {f}^{(0)}_{(z,P)}(\bbp)+  {f}^{(1)}_{(z,P)}(\bbp) \;,
 \end{eqnarray}
 where ${f}^{(1)}_{(z,P)}(\bbp)$ describes the modifications due to the elastic scattering kernel $\bar{\Gamma}(\bbp)$ and satisfies the linear inhomogeneous differential equation
 \begin{align}
 & \left[\partial_{\bbp}^{2} + \frac{3}{\bbp} \partial_{\bbp} 
 - 2 P z(1{-}z) M_{\rm eff}(z,P) 
 \right. \nonumber \\ & \left. \phantom{\frac 12} {}
 +  2 i  P z(1{-}z) \Big[ C_{1} \bar{C}(b) + C_{z} \bar{C}(z \bbp) + C_{1{-}z} \bar{C}((1{-}z)\bbp) \Big]  \right]
 {f}^{(1)}_{(z,P)}(b) 
\nonumber \\ = & -2i  P z(1{-}z) \Big[ C_{1} \bar{C}(\bbp) + C_{z} \bar{C}(z \bbp) + C_{1-z} \bar{C}((1{-}z)\bbp) \Big] {f}^{(0)}_{(z,P)}(\bbp)\;.
 \end{align}
Since the singular part of $ f_{(z,P)}(\bbp)$ is already captured by the contribution ${f}^{(0)}_{(z,P)}(\bbp)$, one concludes that the function ${f}^{(1)}_{(z,P)}(\bbp)$ has to be regular in the vicinity of the origin, i.e.
\begin{eqnarray}
\lim_{\bbp \to 0} {f}^{(1)}_{(z,P)}(\bbp) = c_{1} \,.
\end{eqnarray}
Now as we have already seen from the free case, the above differential equation clearly features diverging solutions in the limit $\bbp \to 0$. Hence the requirement that ${f}^{(1)}_{(z,P)}(\bbp)$ remains finite fixes one of the integration constants in the general solution of the differential equation. Similarly, the second integration constant can be determined from the requirement that the solution remains regular in the limit $\bbp \to \infty$, i.e.
\begin{eqnarray}
\lim_{\bbp \to \infty} {f}^{(1)}_{(z,P)}(\bbp) =0\;.
\end{eqnarray}
Expressing the general solution of the linear inhomogeneous ODE as the general solution to the homogeneous ODE plus a special solution to the inhomogenous ODE and choosing the special solution to be regular for $b \to 0$, we can formally write
\begin{eqnarray}
{f}^{(1)}_{(z,P)}(\bbp)= c_{\rm div} {f}^{(1) {\rm hom, div}}_{(z,P)}(\bbp) + c_{\rm reg} {f}^{(1) {\rm hom, reg}}_{(z,P)}(\bbp) +{f}^{(1) {\rm inhom, reg}}_{(z,P)}(\bbp)
\end{eqnarray}
where the first coefficient $c_{\rm div}$ multiplying the divergent contribution  ${f}^{(1) {\rm hom, div}}_{(z,P)}(\bbp)$ is fixed as $c_{\rm div}=0$ from the requirement that ${f}^{(1)}_{(z,P)}(\bbp)$ is regular. Conversely the second coefficient $c_{\rm reg}$ can be fixed from the requirement that ${f}^{(1)}_{(z,P)}(\bbp)$ remains finite in the limit $\bbp \to \infty$, i.e.
\begin{eqnarray}
\label{eq:cReg}
 c_{\rm reg} = - \lim_{\bbp \to \infty} \frac{{f}^{(1) {\rm inhom, reg}}_{(z,P)}(\bbp)}{{f}^{(1) {\rm hom, reg}}_{(z,P)}(\bbp) }
\end{eqnarray}
Hence, if we choose to solve the differential equations with the following initial conditions in the limit $b\to0$
\begin{align}
{f}^{(1) {\rm hom, reg}}_{(z,P)}(\bbp=0) & =\gs^2 T\;, & {f'}^{(1) {\rm hom, reg}}_{(z,P)}(\bbp=0) & = 0 \;,
\\
{f}^{(1) {\rm inhom, reg}}_{(z,P)}(b=0) & = \gs^2 T\;, &
{f'}^{(1) {\rm inhom, reg}}_{(z,P)}(\bbp=0) & = 0 \;,
\end{align}
we can exploit the fact that ${f}^{(1)}_{(z,P)}(b)$ and its derivatives do not diverge in the limit $b \to 0$ to determine the relevant derivative at the origin ($\nabla_{\bp} \bp = 2$) as
\begin{eqnarray}
 \text{Im}\left[8\nabla_{\bp} \cdot \mathbf{f}_{(z,P)}(\bp) \right]_{\bp=0} = 16 \text{Im}~c_{1} = 16~\gs^2 T~\text{Im}~c_{\rm reg}\;,
\end{eqnarray}
with $ c_{\rm reg}$ given by Eq.~(\ref{eq:cReg}). By combining all the relevant pieces the splitting rate is then determined by
\begin{eqnarray}
\frac{\d \Gamma_{ij}}{\d z} (P,z) =\frac{4}{\pi}~\gs^4 T P_{ij}(z)~\text{Im}~c_{\rm reg} \,.
\end{eqnarray}

\subsection{Numerical implementation}
We solve the homogeneous and inhomogeneous ODE for ${f}^{(1) {\rm hom, reg}}$ and ${f}^{(1) {\rm inhom, reg}}$ numerically using an adaptive fourth order Runge-Kutta scheme. We start at a finite value of $\bbp = \bbp^{\rm min}$, where the initial conditions for ${f}^{(1) {\rm hom, reg}}$ are determined from the series expansion of the solution around $\bbp=0$ as
\begin{eqnarray}
{f}^{(1) {\rm hom, reg}}(\bbp^{\rm min}) &=& \gs^2 T \left( 1 + \frac{1}{4} M_{\rm eff} P z \zb \bbp^{\rm min\; 2} \right) \;,  
\end{eqnarray}
and similarly for the inhomogeneous solution ${f}^{(1) {\rm inhom, reg}}$. Since the latter is sensitive to the elastic scattering kernel, different expansions have to be employed for the different interaction kernels. Specifically, for the LO potential, the small $\bbp$ expansion takes the form
\begin{align}
    {f}^{(1) {\rm inhom, reg}}(\bbp^{\rm min}) & - {f}^{(1) {\rm hom, reg}}(\bbp^{\rm min}) = 
    \frac{-i c(T)}{32\pi}(\gs T \bbp^{\rm min})^2\ln(\gs T \bbp^{\rm min}) \left( C_1 {+} z^2 C_z {+} \bar{z}^2 C_{\bar{z}} \right) \nonumber\\
    &-\frac{i(\gs T \bbp^{\rm min})^2}{128\pi} \left( 
        \left[4 \frac{\hat{q}_0}{4 \gs^4 T^3} + c(T) (4\ln(2)- 3) \right] (C_1+z^2 C_z + \bar{z}^2 C_{\bar{z}}) \right.
        \nonumber\\
        & \left. \phantom{ \frac{i(\gs T \bbp^{\rm min})^2}{128\pi}} 
        \hspace{1.5em} {}+ 4c(T)\left(C_1 \ln\tfrac{1}{2} +z^2 C_z \ln\tfrac{z}{2} + \bar{z}^2 C_{\bar{z}}\ln\tfrac{\bar{z}}{2}\right)
     \right)\;,
\end{align}
where $ c(T) = - \frac{\CR}{8 \pi} \frac{\zeta(3)}{\zeta(2)} \left( -\frac{1}{2 \gs^2} + \frac{3 y}{2} \right) (\gsqb)^2 $.
Based on the terms in the expansion, the starting value $\bbp^{\rm min}$ is then chosen as
\begin{equation}
\bbp^{\rm min}=10^{-4} \text{min}\left(\frac{1}{\sqrt{ M_{\rm eff} P z \zb }},\frac{1}{8 \pi  P z \zb} ,  (\gs T)^{-1}, \frac{1}{\sqrt{(C_{1} +C_{z} z^2+ C_{\zb} \zb^2) (\gs T)^{2}}} \right)\;,
\end{equation}
such that higher order corrections are effectively negligible at double precision level. Starting from the initial conditions at $b=b_{\rm min}$, we then solve the differential equations for ${f}^{(1) {\rm hom, reg}}$ and ${f}^{(1) {\rm inhom, reg}}$, up to a maximal value of $\bbp=\bbp^{\rm max}$, which is chosen similarly according to the criterion that
\begin{equation}
|f^{(0)}(\bbp^{\rm max})/{f}^{(1) {\rm inhom, reg}}(\bbp^{\rm max})| < 10^{-16}\,.
\end{equation}
Based on the solution, we then determine the constant $c_{\rm reg}$ from the ratio 
\begin{equation}
 c_{\rm reg} = - \frac{{f}^{(1) {\rm inhom, reg}}_{(z,P)}(\bbp^{\rm max})}{{f}^{(1) {\rm hom, reg}}_{(z,P)}(\bbp^{\rm max})}\;.
\end{equation}
Such that the differential radiation rate is ultimately given by
\begin{equation}
\frac{\d \Gamma_{ij}}{\d z} (P,z) =\frac{4}{\pi}~\gs^4 T P_{ij}(z)~\text{Im}~c_{\rm reg}\;.
\end{equation}

\section{Bethe-Heitler regime \label{sec:BH}}
We solve Eq.~(\ref{eq:AMY}) perturbatively following an opacity expansion in the number of elastic scatterings, such that at leading order
\begin{align}
\mathbf{g}^{(0)}_{(z,P)}(\ptt) & = \frac{-2i\ptt}{\delta E(z,P,\ptt)} = \frac{-4i \ptt Pz(1{-}z)}{ \ptt^2 + \mu(z)^2}\;, \\ 
\mu(z)^{2} & = (1{-}z) m^2_{\infty,(z)} 
+ zm^2_{\infty,(1{-}z)}-z(1{-}z)m^2_{\infty,(1)}\;,
\end{align}
which is entirely imaginary and thus does not contribute to the splitting rate. Hence the first non-trivial contribution comes from
\begin{align}
2 \ptt \mathbf{g}^{(1)}_{(z,P)}(\ptt)  = \frac{2i \ptt}{\delta E(z,P,\ptt)} \! \int \! \frac{\d ^2\qtt}{(2\pi)^2}~\bar{C}(\qtt)& 
\left\{ C_{1} \left[ \mathbf{g}^{0}_{(z,P)}(\ptt) - \mathbf{g}^{0}_{(z,P)}(\ptt {-}\qtt) \right] \right.  \\
& \left. {} + C_{z} \left[ \mathbf{g}^{0}_{(z,P)}(\ptt) - \mathbf{g}^{0}_{(z,P)}(\ptt {-} z\qtt) \right] \right. \nonumber \\
& \left. {} + C_{1{-}z} \left[ \mathbf{g}^{0}_{(z,P)}(\ptt) -
\mathbf{g}^{0}_{(z,P)}(\ptt {-} \bar{z} \qtt) \right] \right\}. \nonumber
\end{align}
By plugging in the leading order one finds
\begin{align}
2 \ptt \mathbf{g}^{(1)}_{(z,P)}(\ptt) = & 16 P^2 z^2(1{-}z)^2 \int \! \frac{\d ^2\qtt}{(2\pi)^2}~\bar{C}(\qtt)~\frac{1}{\ptt^2+\mu^2(z)}
 \times \nonumber \\
 & \left\{ C_{1}  \left[ \frac{\ptt^2}{\ptt^2+\mu(z)^2} - \frac{\ptt (\ptt-\qtt) }{(\ptt-\qtt)^2+\mu(z)^2}  \right]   \right. \nonumber \\
 & + C_{z}\left[ \frac{\ptt^2}{\ptt^2+\mu(z)^2} - \frac{\ptt (\ptt-z\qtt) }{(\ptt-z\qtt)^2+\mu(z)^2}  \right] \nonumber \\
 & \left. + C_{1-z} \left[ \frac{\ptt^2}{\ptt^2+\mu(z)^2} - \frac{\ptt (\ptt-\bar{z}\qtt) }{(\ptt-\bar{z}\qtt)^2+\mu(z)^2}  \right] \right\}\;,
\end{align}
such that the rate is given by
\begin{align} 
&\frac{\d \Gamma_{ij}^{BH}}{\d z} (P,z) = \gs^4 T  P_{ij}(z) \frac{1}{\pi} \int \! \frac{\d ^2 \ptt}{(2\pi)^2} \int \! \frac{\d ^2\qtt}{(2\pi)^2} \frac{1}{\gs^2 T}~\bar{C}(\qtt) ~\frac{1}{\ptt^2+\mu(z)^2}~\nonumber\\
& \hspace{100pt} \times \left\{ C_{1}  \left[ \frac{\ptt^2}{\ptt^2+\mu(z)^2} - \frac{\ptt (\ptt-\qtt) }{(\ptt-\qtt)^2+\mu(z)^2}  \right]
\right. \nonumber \\
& \hspace{120pt} \left. + \, C_{z}\left[ \frac{\ptt^2}{\ptt^2+\mu(z)^2} - \frac{\ptt (\ptt-z\qtt) }{(\ptt-z\qtt)^2+\mu(z)^2}  \right]  \right. \nonumber \\
& \hspace{120pt} \left. + \, C_{1-z} \left[ \frac{\ptt^2}{\ptt^2+\mu(z)^2} - \frac{\ptt (\ptt-\bar{z}\qtt) }{(\ptt-\bar{z}\qtt)^2+\mu(z)^2}  \right] \right\} \, . 
\end{align}
We perform the re-arrangement
\begin{eqnarray} 
&&\int \! \frac{\d ^2 \ptt}{(2\pi)^2} \frac{1}{\ptt^2+\mu^2(z)}~ \left[ \frac{\ptt^2}{\ptt^2+\mu(z)^2} - \frac{\ptt (\ptt-\qtt) }{(\ptt-\qtt)^2+\mu(z)^2}  \right] \\
&=& \frac{1}{2} \int \! \frac{\d ^2 \ptt}{(2\pi)^2}   \left( \frac{\ptt}{\ptt^2+\mu(z)^2} - \frac{(\ptt-\qtt) }{(\ptt-\qtt)^2+\mu(z)^2}  \right)^{2}.
\end{eqnarray}
To re-write the terms in a manifestly positive definite form, we can re-express the rate as
\begin{eqnarray}
&&\frac{\d \Gamma_{ij}^{BH}}{\d z} (P,z) = \gs^4 T  P_{ij}(z)~Q^{BH}(z,\mDsq,m_{\infty}^2)\;,
\end{eqnarray}
where $Q^{BH}(z,\mDsq,m_{\infty}^2)$ is a dimensionless integral given by
\begin{align}
&Q^{BH}(z,\mDsq,m_{\infty}^2) = \frac{\mDsq}{2\pi \gs^2 T} \int \! \frac{\d ^2 \ptt}{(2\pi)^2} \int \! \frac{\d ^2\qtt}{(2\pi)^2} ~ \bar{C}(\mD \qtt) \nonumber \\
& \hspace{120pt} \times \left\{ C_{1}  \left[ \frac{\ptt}{\ptt^2+\mut(z)^2} - \frac{(\ptt-\qtt) }{(\ptt-\qtt)^2+\mut(z)^2}  \right]^2  \right. \nonumber \\
& \hspace{140pt} \left. + \, C_{z} \left[ \frac{\ptt}{\ptt^2+\mut(z)^2} - \frac{(\ptt-z\qtt) }{(\ptt-z\qtt)^2+\mut(z)^2}  \right]^2 \right. \nonumber \\
& \hspace{140pt} \left. + \, C_{1-z} \left[ \frac{\ptt}{\ptt^2+\mut(z)^2} - \frac{(\ptt-\bar{z}\qtt) }{(\ptt-\bar{z}\qtt)^2+\mut(z)^2}  \right]^2 \right\} \, , 
\end{align}
with  $\mut(z)^{2}=\mu(z)^{2}/\mDsq$. By re-scaling $\ptt$ in the second and third term, the three terms can be expressed in terms of a single integral 
\begin{equation}
 Q^{BH}(z,\mDsq,m_{\infty}^2) = C_{1} Q\Big(\mut^{2}(z)\Big) + C_{z} Q\Big(\frac{\mut^{2}(z)}{z^2}\Big) + C_{z} Q\Big(\frac{\mut^{2}(z)}{(1{-}z)^2}\Big)\;,
\end{equation}
where
\begin{eqnarray}\label{eq:QIntegral}
Q(\mut^2)=  \frac{\mDsq}{2\pi \gs^2 T} \int \! \frac{\d^2 \ptt}{(2\pi)^2} \int \frac{\d ^2\qtt}{(2\pi)^2} ~ \bar{C}(\mD \qtt)  \left[ \frac{\ptt}{\ptt^2+\mut^2} - \frac{(\ptt-\qtt) }{(\ptt-\qtt)^2+\mut^2}  \right]^2\;.
\end{eqnarray}
Now evaluating $\mut(z)^{2}$ for the different channels, we get
\begin{eqnarray}
&&\mut^{2}_{g\to gg}(z)= (1{-}z) \frac{m^2_{\infty,(g)}}{\mDsq}+z\frac{m^2_{\infty,(g)}}{\mDsq}-z(1{-}z)\frac{m^2_{\infty,(g)}}{\mDsq} = \frac{1-z(1{-}z)}{2}\;,\\
&&\mut^{2}_{q\to gq}(z)=(1{-}z) \frac{m^2_{\infty,(g)}}{\mDsq}+z\frac{m^2_{\infty,(q)}}{\mDsq}-z(1{-}z)\frac{m^2_{\infty,(q)}}{\mDsq}=\frac{1-z}{2}+az^2\;,\\
&&\mut^{2}_{g\to qq}(z)= (1{-}z) \frac{m^2_{\infty,(q)}}{\mDsq}+z\frac{m^2_{\infty,(q)}}{\mDsq}-z(1{-}z)\frac{m^2_{\infty,(g)}}{\mDsq}= \frac{2a-z(1{-}z)}{2}\;,
\end{eqnarray}
where $a=m^2_{\infty,q}/\mDsq$, such that the rates can be compactly expressed as
 \begin{align}
 &\frac{\d \Gamma_{g\to gg}^{BH}}{\d z} (P,z) =
 \gs^4  T P_{g\to gg}(z) \times \\
 & \left[  \frac{C_A}{2} Q\left( \frac{1-z(1{-}z)}{2} \right)  + \frac{\CA}{2} Q\left( \frac{1-z(1{-}z)}{2z^2} \right) + \frac{\CA}{2} Q\left( \frac{1-z(1{-}z)}{2(1{-}z)^2} \right) \right]\;, \nonumber \\ 
&\frac{d\Gamma_{q\to gq}^{BH}}{dz} (P,z) =
 \gs^4 T P_{q\to gq}(z) \times \nonumber \\
 & \left[ \frac{\CA}{2} Q\left( \frac{1-z+2a z^2}{2} \right)  
 + \left(\CF - \frac{\CA}{2} \right) Q\left( \frac{1-z+2az^2}{2z^2} \right) +\frac{\CA}{2} Q\left( \frac{1-z+2az^2}{2(1{-}z)^2} \right) \right] \;, \nonumber \\ 
&\frac{d\Gamma_{g\to qq}^{BH}}{dz} (P,z) =
 \gs^4 T  P_{g\to qq}(z) \times \nonumber \\
 & \left[ \left(\CF - \frac{\CA}{2} \right) Q\left( \frac{2a-z(1{-}z)}{2} \right)  +\frac{\CA}{2} Q\left( \frac{2a-z(1{-}z)}{2z^2} \right) +\frac{\CA}{2} Q\left( \frac{2a-z(1{-}z)}{2(1{-}z)^2} \right) \right]\;.\nonumber
\end{align}

\subsection{Evaluating the integral in impact-parameter space}
In this section we show how to compute the $Q(\tilde\mu^2)$ integral in $\bp$-space. We start by rewriting the integral in Eq.~(\ref{eq:QIntegral}) as follows:
\begin{equation}
Q(\tilde\mu^2) = \frac{1}{\gs^2 T} \int \! \frac{\d ^2 \ptt}{(2\pi)^2} ~ \vec{\psi}(\ptt) \int \! \frac{\d ^2\qtt}{(2\pi)^2}~\bar{C}(\qtt)
 \left( \vec{\psi}(\ptt)- \vec{\psi}( \ptt - \qtt) \right)\;,
\end{equation}
where we introduce the function $\vec{\psi}(\ptt) =\frac{\ptt}{\ptt^2+\tilde\mu^2}$. 
Its Fourier transform is given by
\begin{equation}
\vec{\psi}(\bp) = \int \! \frac{\d ^2\qtt}{(2\pi)^2} e^{-i\qtt\cdot\bp} \vec{\psi}(\qtt)=-\frac{i}{2\pi}\tilde\mu K_1(\tilde\mu \bbp) \frac{\bp}{\bbp} \, ,
\end{equation}
where $K_1(x)$ is the modified Bessel function of the second kind.
Inserting the Fourier transform to the $Q(\tilde\mu^2)$ integral and using the definition of the broadening kernel in Eq.~(\ref{subtraction_FT}) we obtain
\begin{align}
Q(\tilde\mu^2) &=  \frac{1}{\gs^2 T} \int \! \d ^2\bp \!\! \int \! \frac{\d ^2 \ptt}{(2\pi)^2} ~ \vec{\psi}(\ptt) \!\int  \frac{\d ^2\qtt}{(2\pi)^2} ~\bar{C}(\qtt) e^{i\ptt\cdot\bp}\left( 1 - e^{-i\bp\cdot\qtt} \right)\vec{\psi}(\bp)  \nonumber \\
&= \frac{1}{\gs^2 T} \int \! \d ^2\bp \int \! \frac{\d ^2 \ptt}{(2\pi)^2} ~ \vec{\psi}(\ptt)~\bar{C}(\bp) e^{i\ptt\cdot\bp}\vec{\psi}(\bp)
\nonumber \\
&=\frac{1}{\gs^2 T} \int \! \d ^2 \bp ~\bar{C}(\bp)\vec{\psi}(-\bp)\cdot\vec{\psi}(\bp) \nonumber \\
&= \frac{1}{\gs^2 T} \int_{0}^{\infty} \! \frac{\d \bbp}{2\pi} ~\bar{C}(\bbp) \, \bbp \tilde\mu^2 K_1(\tilde\mu \bbp)^2\;.
\end{align}
This last integral is equivalent to Eq.~(\ref{eq:QIntegral}). However, being in position space, we can use it to obtain the rate in the Bethe-Heitler regime for the non-perturbative kernel.

\FloatBarrier

\bibliographystyle{ieeetr}
\bibliography{biblio}

\end{document}